# Magnetic properties of thin epitaxial $Pd_{1-x}Fe_x$ alloy films

A. Esmaeili[1], I.V. Yanilkin[1], A.I. Gumarov[1], I.R. Vakhitov[1], B.F. Gabbasov[1],

D.A. Tatarsky[2], R.V. Yusupov[1], L.R. Tagirov[1,3]

[1]Kazan Federal University, Kremlevskaya18, Kazan 420008, Russia

[2]Institute for Physics of Microstructures of RAS, GSP-105, Nizhny Novgorod 603950, Russia

[3]Zavoisky Physical-Technical Institute, FRC "Kazan Scientific Center of RAS", Sibirsky Trakt, 10/7, Kazan 420029, Russia

Corresponding author E-mail: yanilkin-igor@yandex.ru. Mobile phone: +79872081176.

In the paper we present the results of extensive studies of palladium-rich $Pd_{1-x}Fe_x$ alloy films epitaxially grown on MgO single-crystal substrate. In a composition range of $x = 0.01$-$0.07$ these materials are soft ferromagnets, the saturation magnetization and magnetic anisotropy of which can be tuned by its composition. Vibrating sample magnetometry was used to study temperature dependences of spontaneous magnetic moment and to establish the temperature of magnetic ordering (Curie temperature). Ferromagnetic resonance (FMR) measurements at low temperatures in the in-plane and out-of-plane geometries revealed the four-fold in-plane magnetic anisotropy with the easy directions along the ⟨110⟩ axes of the substrate. The modelling of the angular dependence of the field for resonance allowed to extract the cubic and tetragonal contributions to the magnetic anisotropy of the films and establish their dependence on the concentration of iron in the alloy. Experimental data are discussed in the framework of existing theories of dilute magnetic alloys. Using the anisotropy constants established from FMR, the magnetic hysteresis loops are reproduced utilizing the Stoner-Wohlfarth model thus indicating the predominant coherent magnetic moment rotation at low temperatures. The obtained results compile a database of magnetic properties of a palladium-iron alloy considered as a material for superconducting spintronics.

## 1. Introduction

Since works by Gerstenberg [1], Crangle *et al.* [2,3] and Clogston *et al.* [4] magnetic properties of palladium-iron ($Pd_{1-x}Fe_x$) alloys have been intensively studied because of unusually low concentration of iron (below 1 at.%) sufficient to induce ferromagnetism (FM) at low temperatures, and giant magnetic moment, up to 12 Bohr magnetons per iron atom in the FM state. Detailed investigations summarized in the early review [5] and numerous textbooks (see for example, [6-10]) have led to understanding that ferromagnetism in palladium-iron alloys (also applies to some other dilute palladium/platinum – $3d$-transition metal alloys) establishes as a result of covalent polarization of itinerant palladium $4d$-electrons around a $3d$-metal dopant ion [8-10]. Once the spin-polarized bubbles join via percolation into a macroscopic cluster, the long-range ferromagnetic order establishes.

The palladium-rich $Pd_{1-x}Fe_x$ alloys have attracted an emergent attention because of potential applications in superconducting spintronics based on Josephson π-junctions [11-15]. In the latter, FM material is used for a thin-film weak link between superconducting terminals and shifts the phase of the superconducting pairing wave-function by π [16]. The superconducting coherence length in the FM layer must be much larger than the technological scatter of its thickness and roughness of its interfaces to make the ferromagnetic phase invertor properties insensitive to the layer inhomogeneities. At the same time, the FM layer should accommodate half of the pairing function oscillation length across the thickness. This leads to the typical length scale of 10-30 nm for both the coherence length and the ferromagnetic layer thickness. None elemental or binary ferromagnetic compound satisfies these requirements. For example, the superconducting coherence length of metallic iron was estimated as $\xi_S(\text{Fe}) \approx 0.7$ nm [17,18], and for metallic nickel as $\xi_S(\text{Ni}) \approx 0.9$ nm [19]. So short coherence lengths are a consequence of high Curie temperatures (1043 K and 627 K for iron and nickel, respectively) and large exchange splitting energies of the conduction band in strong ferromagnets (a fraction of the electron-volt, see review [20] for further details).

In the context of the above, palladium-rich $Pd_{1-x}Fe_x$ alloy looks promising due to a known possibility to tune its magnetic properties by varying the iron concentration $x$ and preparation conditions. Indeed, the $T_c$ value for $x = 0.01 – 0.1$ varies in the range of 30 – 220 K [5] thus offering a room to adjust the coherence length within an order of its magnitude. This, however, refers to bulk alloys; much less works were devoted to thin palladium-iron films. In particular, predominantly the polycrystalline films with low concentration of iron have been investigated [15, 21-25]. We could find only few reports on studies of epitaxial Pd-rich films [26,27] (and possibly Refs. [28,29], albeit no evidences of the epitaxial film growth had been given there). Recently, we have successfully adopted an advanced three-step procedure [30] for a synthesis of high-quality epitaxial $Pd_{1-x}Fe_x$ ($x < 0.10$) thin films on (001)-oriented MgO substrates [31]. The "cube-on-cube" epitaxy was realized for those films. The principal characteristics that determine an applicability of $Pd_{1-x}Fe_x$ alloys for ultrathin-layer S/F/S-type structures are the saturation magnetization and magnetic anisotropy. The latter can be effectively probed with ferromagnetic resonance spectroscopy. In this paper we present a systematic study of the magnetic and magnetoresonant properties of $Pd_{1-x}Fe_x$ ($x = 0.01-0.07$) alloy thin films epitaxially grown on the MgO substrate, targeting a low-temperature weak and tunable ferromagnet suitable for cryogenic spintronics.

2. **Experimental techniques**

Epitaxial films of $Pd_{1-x}Fe_x$ ($x = 0.011 – 0.07$) alloy with the thickness of 20 nm were grown on epi-polished (100)-oriented MgO substrates (*MTI Corp., USA*) utilizing molecular beam epitaxy (MBE) technique in the ultra-high vacuum (UHV) chamber (*SPECS GmbH, Germany*). The pressure during the depositions was in the range of $(3–5)\times10^{-10}$ mbar. The $^{57}$Fe isotope (enrichment of 98.5% for the samples to be suitable for Mössbauer spectroscopy measurements) metallic pellets and Pd metal pellets were separately co-evaporated from high-temperature effusion cells (*CreaTec, Germany*). Three-step deposition procedure [30,31] was applied to obtain 20 nm thick, smooth and compositionally homogeneous epitaxial films of $Pd_{1-x}Fe_x$ alloy. The representative transmission electron microscopy (TEM) image and the selected-area electron diffraction pattern (SAED) from the MgO/$Pd_{1-x}Fe_x$ interface (Fig. 1) confirm epitaxial growth of the palladium-iron alloy film on the single-crystalline (100)-MgO substrate. Details of the samples preparation and characterization routines can be found in Ref. [31]. Magnetic properties of the synthesized samples were studied with vibrating sample magnetometry (VSM, *Quantum Design PPMS-9 system*) and ferromagnetic resonance (FMR, *Bruker ESP300 cw X-band spectrometer*) techniques. The FMR measurement and interpretation routines are described in Ref. [32].

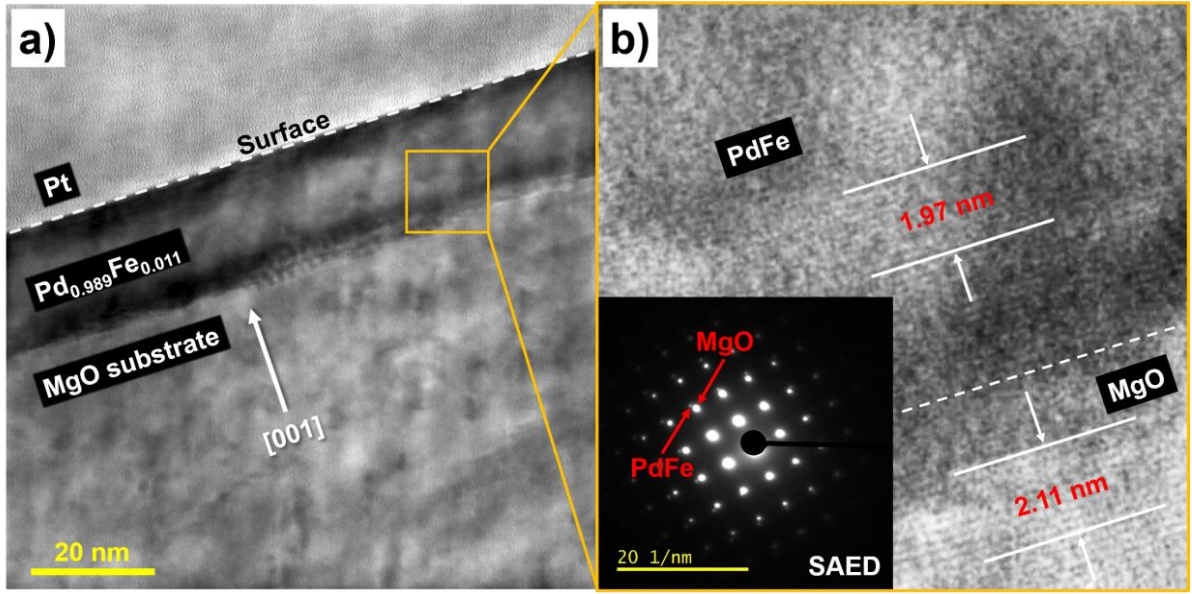

Figure 1. a) TEM image of the sample cross-section; b) magnified area of the $Pd_{0.989}Fe_{0.011}$/MgO(001) interface where ten interplanar distances are marked for both materials. The inset shows the selected area electron diffraction (SAED) pattern taken from the interface. High quality of the contrast, reflecting the cubic symmetry of the lattice, is an indication of the single-crystal character of the interface area. Two sets of the superimposed diffraction spots correspond to the MgO lattice (smaller pitch size because of larger lattice parameter $a_{MgO}$=0.4216 nm) and $Pd_{0.988}Fe_{0.012}$ lattice (larger pitch size because of smaller lattice parameter $a_{Pd}$=0.3891 nm).

The temperature dependences of the magnetic moment $M(T)$ were measured on sample warming from 5 K to 300 K at a rate of 3 K/min in the magnetic field of 30 Oe applied in the film plane ("in-plane" geometry) along the film easy axis. Magnetic hysteresis loops were recorded with the sweep rate of 2.3 Oe/sec. The measured magnetic moment was reduced to the number of Bohr magnetons ($\mu_B$) per iron atom as well as to the volume magnetization of the film (emu/cm$^3$). The diamagnetic and paramagnetic contributions of the MgO substrate were subtracted from the collected VSM data.

### 3. Experimental results

Figure 2a shows the temperature dependence of the saturation magnetization for a composition series of $Pd_{1-x}Fe_x$ films. In Fig. 2b, the same set of the dependencies is presented in the reduced $M_s/M_s(0)$ vs $T/T_C$ coordinates. Clearly, these temperature dependencies have practically identical shape for all studied concentrations of iron, though this common dependence is significantly different from that for the pure iron (also shown in Fig. 2b). We will address this observation later, in the Discussion section.

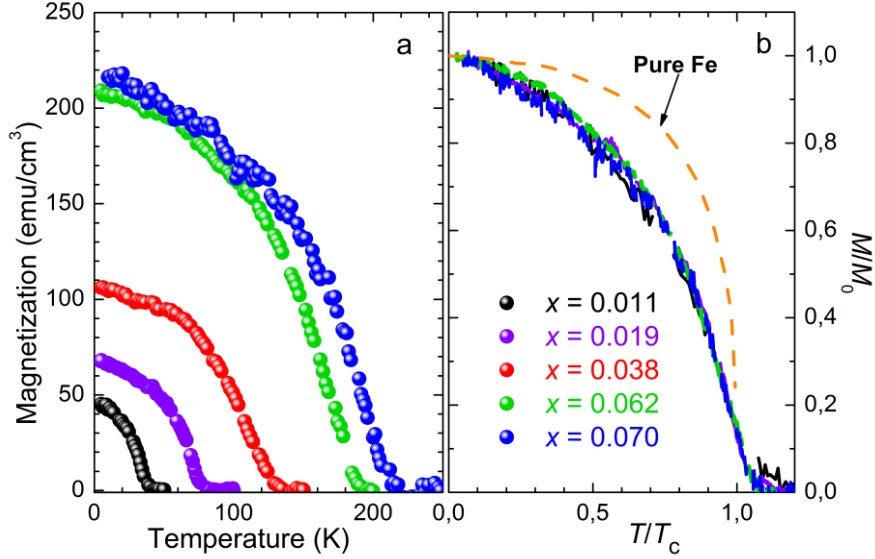

Figure 2. Temperature dependence of the saturation magnetization for a series of 20-nm thick epitaxial $Pd_{1-x}Fe_x$ films on (001)-MgO, in the absolute (a) and the reduced (b) coordinates. Normalized $M_s(T/T_C)/M_s(0)$ dependence for pure iron is given for a comparison (orange symbols, adopted from Ref. [3]).

As expected, both the film saturation magnetization $M_s$ and the Curie temperature $T_C$ raise monotonically with the increase of the iron content $x$ (see Table 1). Maximum magnetic moment per iron atom of $M_{Fe} \simeq 7.5\ \mu_B$/Fe is found for the $x = 0.017$ sample. This is in a good agreement with the results for the bulk $Pd_{1-x}Fe_x$, where the values of $\simeq 10\ \mu_B$/Fe for $x = 0.01$ and $\simeq 5.5\ \mu_B$/Fe for $x = 0.1$ have been reported [3]. However, the obtained result is considerably larger than ~ 3.7 $\mu_B$/Fe for a non-epitaxial $Pd_{0.99}Fe_{0.01}$ film investigated in [24].

Table 1. Experimental parameters of epitaxial $Pd_{1-x}Fe_x$ alloy films with different iron concentrations.

| $x$, from XPS | $T_C$ (K) | $M_S$ (emu/cm³) | *$M_{Fe}$ ($\mu_B$) | **$H_c$ (Oe) | | ***g-factor |
|---|---|---|---|---|---|---|
| | | | | easy | hard | |
| 0.011±0.001 | 36±2 | 66±4 | 6.8±0.9 | 8.8±1 | 7±1 | 2.19 |
| 0.014±0.001 | 49±2 | 63±4 | 7.1±0.9 | 7.5±1 | 5.5±1 | 2.2 |
| 0.017±0.001 | 72±3 | 83±5 | 7.5±0.8 | 16.7±1 | 14.7±1 | 2.18 |
| 0.019±0.002 | 73±3 | 85±5 | 6.5±0.8 | 10.1±1 | 7.8±1 | 2.15 |
| 0.038±0.002 | 120±4 | 119±6 | 5.3±0.6 | 8.5±1 | 9±1 | 2.15 |
| 0.062±0.002 | 177±5 | 211±11 | 5.6±0.5 | 17±1 | 18.5±1 | 2.17 |
| 0.070±0.002 | 200±5 | 216±11 | 4.3±0.3 | 23±1 | 15±1 | 2.13 |

* Magnetic moment per iron atom.
** Coercive filed measured at 5 K along easy or hard axes of the film.
***For g-factor experimental errors are less than 0.004 for all samples.

Magnetic anisotropy of the films has been characterized utilizing the FMR technique. The FMR spectra were recorded at $T = 10$ K for the samples with $x = 0.011 – 0.019$ ($T_C < 100$ K), and at $T = 20$ K for the samples with higher iron content. Figure 3 shows the representative angular dependences of the FMR resonance field for the in-plane geometry of the measurements (external magnetic field is rotated in the film plane). The samples reveal the four-fold in-plane magnetic anisotropy with the easy directions along the $\langle 110 \rangle$ axes, which is typical for these materials. For instance, Bagguley *et al.* [33] reported that the bulk single-crystalline $Pd_{0.983}Fe_{0.017}$ alloy reveals the cubic anisotropy with the hard directions along the $\langle 100 \rangle$ axes [33]. The four-fold anisotropy with the easy $\langle 110 \rangle$ axes in the film plane was found also for epitaxial $Pd_{1-x}Fe_x$ ($x \simeq 0.04$ and $0.10$) films by Garifullin *et al.* [27].

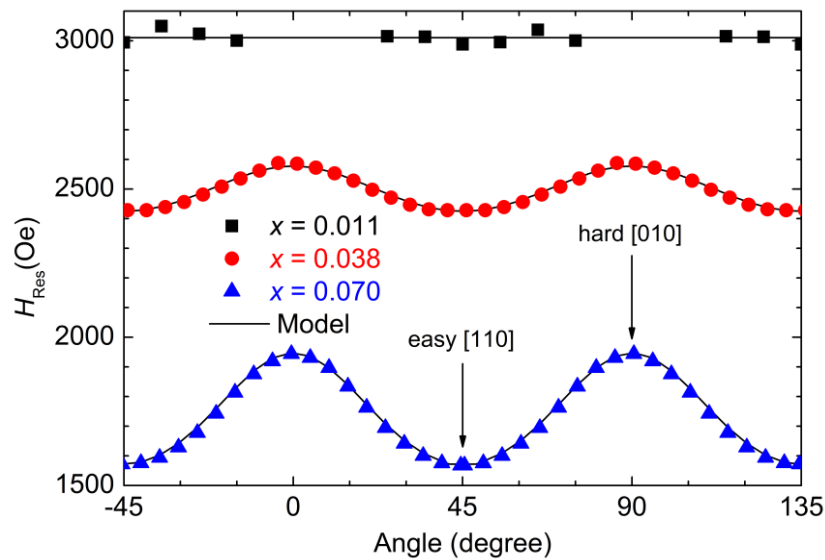

Figure 3. Angular dependences of the FMR resonance fields for the in-plane measurements of $Pd_{1-x}Fe_x$ films with different iron concentrations (symbols – experimental data, lines – fitting results); $T = 10$ K for $x = 0.011$ sample and $T = 20$ K for $x = 0.036$ and $x = 0.07$.

Figure 4 shows hysteresis loops for three representative $Pd_{1-x}Fe_x$ films with different iron concentrations. It is clear from the shape of the loops, as well as from the FMR angular dependences, that the anisotropy becomes stronger with an increase of the iron content $x$ (see Figs. 3 and 4). Coercivity $H_c$ also depends on the iron concentration and varies from 5.5 to 23 Oe (see Table 1). The obtained coercive fields are consistent with those reported for polycrystalline films, compare with $H_c \simeq 5$ Oe for $x = 0.01$ [25] and $H_c \simeq 25$ Oe for $x = 0.072$ [29].

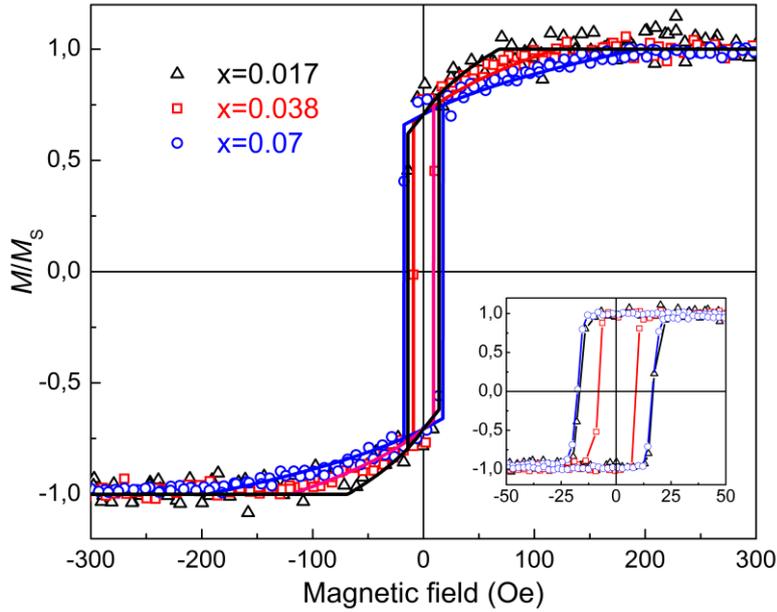

Figure 4. In-plane magnetic hysteresis loops of $Pd_{1-x}Fe_x$ alloy films measured at 5 K along the hard [100] axis of the film. Inset − the same along the easy [110] axis.

## 4. Discussion

To describe the temperature dependence of the magnetic moment and the concentration dependence of the ferromagnetic transition temperature $T_C$ several theoretical models can be applied. These models utilize different approaches in their theoretical frameworks and provide physical parameters suitable for a comparative analysis.

### 4.1 Spin-fluctuation model

The magnetic moment per iron atom in $Pd_{1-x}Fe_x$ series (Table 1) varies from 4.3 $\mu_B$ to 7.5 $\mu_B$ at saturation, while that in metallic iron is about 2.2 $\mu_B$ (see, for example, Ref. [8], Ch. 5, Table 5.1; and Ref. [10], Ch. 5, Table 5.4). This suggests a dominance of the palladium contribution to the total magnetic moment of $Pd_{1-x}Fe_x$ ($x = 0.011 – 0.07$) alloy (or almost equivalent contributions from iron and palladium in the case of the most concentrated sample, $x = 0.07$, with 4.3 $\mu_B$ per iron). The universal temperature dependence of the magnetic moment, Fig. 2b, for different concentrations of iron in the studied range makes it reasonable to fit the $M(T)$ data within the most general approach, which takes into account the itineracy of electrons carrying magnetism. The approach is based on the Landau-type expansion on powers of magnetization to higher orders to account for spin fluctuations [8, 34-38]. In the latest analysis by Kuz'min [38], an equation for the reduced magnetization $\sigma(T) = M_s(T)/M_s(0)$ (at zero magnetic field) is given, which can be used to fit the experimental $M_s(T)$ from Fig. 2b:

$$\sigma^2(\tau) = \frac{\sqrt{\kappa^2 + 4(1-\kappa)(1-\tau^3)/(1+p\tau^{3/2})} - \kappa}{2(1-\kappa)} \quad (1)$$

Here, $\tau = T/T_C$, and $\kappa$ is a dimensionless ratio of the Landau expansion coefficients (of the fourth order to the second order), which depends on the relationship of nonlinear and linear susceptibilities balanced by the saturation magnetization extrapolated to zero temperature [38]. The parameter $p$ provides correct low-temperature behavior of the magnetic moment $M(T)$ (for example, providing compliance with the Bloch-$T^{3/2}$ law for fixed-spin Heisenberg model with spin-wave excitations). The fitting of the experimental data, see in Fig. 5, has given $p = -0.02 \pm 0.05$, $\kappa = 2.1 \pm 0.1$. The parameter set is in drastic contrast with that of $p = 0.95 \pm 0.12$, and $\kappa = 0.86 \pm 0.07$, obtained for gadolinium [39], which is the localized-spin, $S = 7/2$ ferromagnet. Similar data for nickel, which is the iron-group ferromagnetic metal with much higher degree of delocalization of its $d$-electrons compared with $f$-electrons of gadolinium, has given $p = 0.28$ and $\kappa = 0.47$ [38] which lie somewhere in between $Pd_{1-x}Fe_x$ and gadolinium. This observation provides a further evidence of a dominant itinerant carriers contribution to the spontaneous magnetization of $Pd_{1-x}Fe_x$ alloys in the range of $x = 0.01 - 0.07$.

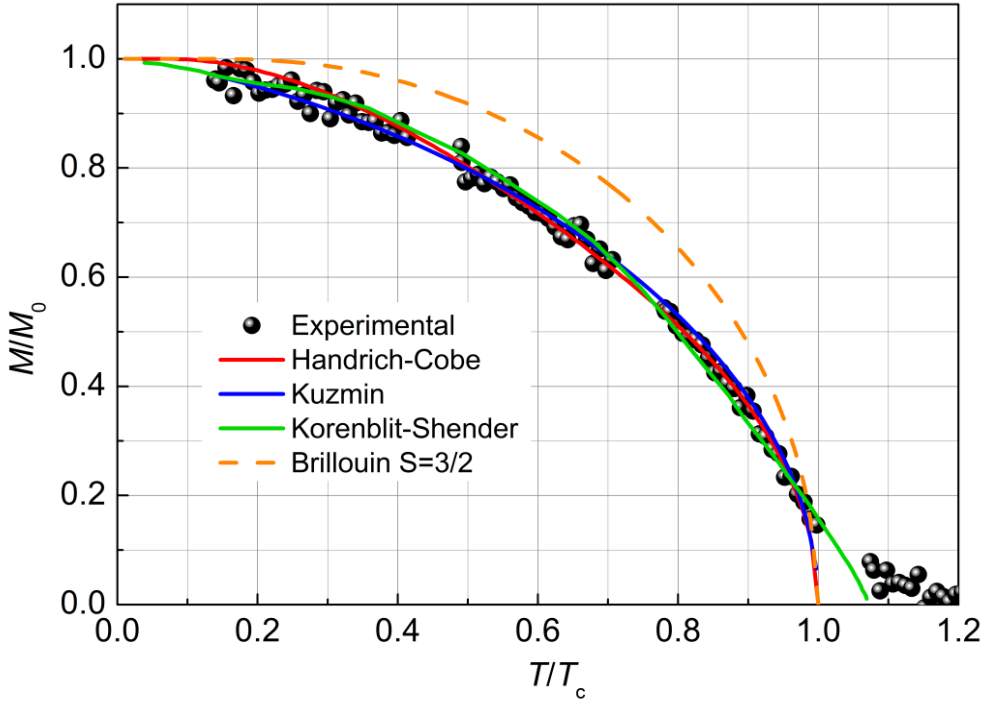

Figure 5. Normalized saturation magnetization $M_s/M_s(0)$ dependence on the reduced temperature $T/T_C$ of the $Pd_{0.989}Fe_{0.011}$ film (solid symbol) and its fits with three theoretical models (lines), see the legend and main text.

*4.2 Handrich-Kobe model*

Nanoscale composition inhomogeneity is one of the main differences between alloys and elementary metals or stoichiometric compounds. This feature also affects the shape of the magnetization dependence on temperature. Handrich-Kobe model [40,41] uses a modified Brillouin function with the distribution of the exchange coupling strength in the positionally disordered environment. The reduced magnetization is expressed as:

$$\sigma(\tau) = \frac{1}{2}\{B_S[(1+\delta)x] + B_S[(1-\delta)x]\}, \qquad (2)$$

where $x = \frac{3S}{S+1}\frac{\sigma(\tau)}{\tau}$, $B_S$ —the conventional (spin-only) Brillouin function and $S$ is the magnetic atom spin value. The exchange fluctuation parameter is defined as $\delta = \sqrt{\langle\Delta J^2\rangle/\langle J\rangle^2}$, where $J$ is the exchange integral.

Taking $S = 3/2$, as it follows from the closest proximity from the above to provide the magnetic moment value per atom $\mu$(Fe) = 2.2 $\mu_B$ in metallic iron, we obtain $\delta = 0.48$ from the fit of our experimental data (see Fig. 5). The sensitivity of the theory to the spin value appears to be weak: $\delta$-value varies in the range of 0.45 – 0.53 if the spin $S$ value is varied from 5/2 to 1, respectively. The fluctuations of the exchange integral flatten the $M(T)$-dependence against the ideal structural order ($\delta = 0$ for pure iron) (see Fig. 5). The value of $\delta \approx 0.48$ is close to that found for amorphous iron-containing ferromagnets, for instance, $\delta = 0.5$ for $Fe_{80}B_{12}Nb_8$ [39] and $\delta = 0.55$ for $Fe_{88}Zr_7B_4Cu_1$ [42]. Exchange fluctuation parameter $\delta$ originates from the structural disorder and corresponding distribution of distances to the nearest magnetic neighbors. In this respect, $Pd_{1-x}Fe_x$ alloy with random arrangement of iron atoms is just another nonzero-$\delta$ case. Hence, the application of the Handrich-Kobe approach seems to be adequate, despite it completely ignores the contribution of the palladium constituent to the net moment of the alloy and independence of the $M(T)$ shape on iron concentration in the $Pd_{1-x}Fe_x$ alloy.

*4.3 Korenblit-Shender model*

A commonly accepted physical picture supposes that an iron atom embedded into paramagnetic palladium magnetizes electrons of the host matrix around it because of the *d*-electron hybridization (see, for example, Ref. [8], Ch. 15). Thus, it creates a local "bubble" of magnetized palladium atoms around. On cooling, the bubbles of polarized palladium electrons grow in size, and at a temperature when they start to overlap, the long-range ferromagnetic order sets in. The onset of ferromagnetism thereby occurs through a percolation process involving these spin-polarized bubbles (see, for example, estimates in Refs. [43,44]). Korenblit and Shender [45] have elaborated a model of magnetic ordering for Pd-Fe like alloys with low concentration of ferromagnetism-inducing dopant assuming a development of magnetically-polarized clouds (bubbles) around them. The proposed potential of an indirect exchange interaction between two ferromagnetic atoms had a form:

$$V(r) = V_0 \frac{R}{r} e^{-r/R}. \tag{3}$$

Here $V_0$ determines the interaction energy scale, $R = a/\sqrt{1+\Gamma} \gg a$ is a range of the interaction, where $a$ is the interatomic distance, and $(1+\Gamma)^{-1}$ is the static susceptibility enhancement factor of the host. Magnetic moments of the iron impurity atoms are considered ferromagnetically aligned if the distance between them is less than the critical value of $r_c(T) = R \ln(V_0/k_B T)$. The reduced magnetization $\sigma$ at a certain temperature $T$ is equal to a probability of a magnetic ion occurrence inside the "infinite" magnetic cluster uninterruptedly extending over the entire sample. We have calculated $\sigma(T)$ using the results of the Monte-Carlo simulations in Ref. [46] in combination with the temperature-dependent critical distance $r_c(T)$ [45]. The resulting $\sigma(T)$ dependence can have a look of a flattened convex curve for $(4/3)\pi n R^3 \sim 10^{-2} \div 10^{-3}$ or even a concave curve for $(4/3)\pi n R^3 < 10^{-4}$ with the threshold depending on the dopant concentration and the actual range of interaction provided by the host ($n$ is a number of the dopant atoms per unit volume). The "tail" above the $\sigma(T \to T_c) \to 0$ cannot be described by this "mean-distance" theory because it does not account for the short-range order effects of nearest or close neighbors.

The result of the experimental $M(T)$ data fit with the Korenblit-Shender theory is presented in Fig. 5 for the sample of Pd$_{1-x}$Fe$_x$ alloy with $x=0.011$ (recall that according to Fig. 2b the reduced $\sigma(\tau)$ curves have almost identical shape for $x = 0.011 - 0.07$). The best-fit parameter $R$ varies from $\simeq 1.3$ Å to $\simeq 0.7$ Å for iron content within the range from 0.011 to 0.07. These values look unexpectedly small and concentration dependent, the latter has not been inherent to the original theory [45].

Figure 5 presents the fits with the three models mentioned above to the experimental $M(T)$ data for the Pd$_{1-x}$Fe$_x$ alloy film with the iron content $x = 0.011$. All the three models fit the data fairly well; however, a selection of the best fit basing on the formal minimum of the mean square deviations ($\chi^2$) would be incorrect because of very different assumptions put into the basis of these theories. Additional arguments could be obtained from the concentration dependence of the transition temperature $T_C(x)$.

*4.4 Concentration dependence of the ferromagnetic transition temperature*

The spin-fluctuation model of itinerant electron paramagnets doped by magnetic impurities was developed by Mohn and Schwartz to calculate the temperature of the ferromagnetic transition in alloys of palladium with elements of the iron group [8,47]. The transition temperature $T_C$ is determined by the divergence point of the magnetic susceptibility, which, in turn, is given by the balance of the inverse susceptibilities of the host and magnetic impurity contributions. The concentration-dependent $T_C$ can be found then as a real-valued solution of the cubic equation,

$$A + 5B\alpha k_B T_C + 35C(\alpha k_B T_C)^2 = \frac{(g-1)^2}{\mu_B^2 n} I^2 \frac{2S(S+1)}{3k_B T_C}, \quad (4)$$

where $A$, $B$ and $C$ are the expansion coefficients of the Landau theory, $\alpha$ is the proportionality coefficient between the mean square of the fluctuating magnetic moment of the host and temperature (see [47] and references therein), $g$ and $S$ are the spectroscopic $g$-factor and spin of magnetic impurity (iron in our case), and finally, $I$ is the constant of coupling between the localized impurity moments and the itinerant electron spin. Left-hand side of Eq. (4) is nothing but the inverse magnetic susceptibility of palladium, the fitting of which to a pure palladium susceptibility data fixes parameters $B\alpha/A$ and $C\alpha^2/A$, namely, $5B\alpha k_B/A = 172*(35C(\alpha k_B)^2/A)$ K$^{-1}$ from the maximum of Pd susceptibility at $T_M = 86$K; and $(35C(\alpha k_B)^2/A) = 1.4\times10^{-5}$ K$^2$ from $\chi(T_M)/\chi(T=0) \approx 1.115$. All poorly determined quantities in the numerator of the r.h.s. (Eq.(4)) can be absorbed into a single parameter being the subject to the fitting procedure: (r.h.s./A) ≈ 2920 K. The resulting concentration dependence of $T_C$ is found non-linear because of the non-monotonous temperature dependence of the palladium magnetic susceptibility (Ref. [5], Fig. 2.1) and hence, the l.h.s. of Eq. (4). At low $x$-values, the $T_C$ of Pd$_{1-x}$Fe$_x$ grows with increasing the magnetic dopant concentration; the host susceptibility rises too until it passes through the maximum at $T \sim 86$K. With further increase of the dopant concentration, the growing $T_C$ meets with the falling temperature dependence (Curie regime) of the host susceptibility [5]. As a result, the concentration dependence $T_C(x)$ gradually switches from higher increment at low concentrations to the lower increment once $T_C$ surpasses ~ 86K. The dopant content threshold is within 0.03 – 0.045 of iron in palladium (see Fig. 6b).

Korenblit and Shender suggested an expression for the $T_C$ dependence on $x$ within their model of dilute alloy [45]. The Curie temperature corresponds to a percolation of spin-polarized bubbles and occurrence of the "infinite" magnetic cluster:

$$T_C = V_c \frac{Rn^{1/3}}{0.87} \exp(-\frac{0.87}{Rn^{1/3}}), \tag{5}$$

where $V_c \approx V_0 S^2$, and the numerical factor comes from the percolation theory. The direct fitting of Eq. (5) to the experimental data (see Fig. 6) allows to determine the pre-factor $V_c$ in Eq.(5) as $V_c = 650 \pm 120$ K, and the $R$ value of $4.7 \pm 0.4$ Å. The $T_C(x)$ dependence is non-linear convex, and this non-linearity is ascribed in Ref. [45] to spatial fluctuations of molecular fields, especially at the lower side of concentrations. The fitting is quite good, however, the mismatch between $R \sim 1$ Å and $R \sim 4.7$ Å, determined from the fitting to $M_s(T)$ and $T_C(x)$ data, respectively, looks confusing with no explanation at the moment.

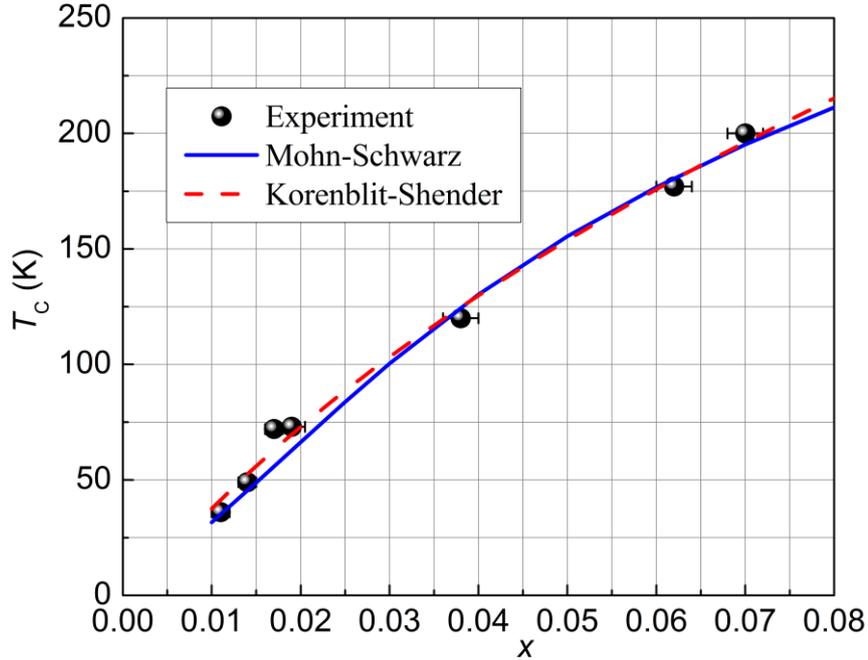

Figure 6. Curie temperature dependence on the iron content $x$ (solid symbols) in Pd$_{1-x}$Fe$_x$ alloy films. Solid blue line is a fit with the Mohn-Schwarz model; and dashed red line is a fit with the Korenblit-Shender model.

Both models describe the concentration dependence of the magnetic transition temperature quite well. However, the Korenblit-Shender model [45] requires the effective range of the interaction between the impurity adatoms $R \gg a$, where $a$ is interatomic distance (see Eq.(3) and its description), whereas fitting to our experimental data gives for the $R$ parameter maximal value of 4.7 Å which is about two interatomic distances. Moreover, the Korenblit-Shender model predicts gradual change of the $M_s(T)$-dependence shape from convex on the high impurity concentration side towards the concave upon dilution. No indication of this trend is visible in our experimental data on $M_s(T)$, Fig. 5. On the other hand, the Mohn-Schwarz spin-fluctuation model does not imply the $M_s(T)$ shape to change until the host susceptibility dominates in the net magnetic response of the alloy. It seems (see the fourth column of Table 1 above) that for our samples with the iron content in the range $x = 0.011 – 0.062$ the palladium host magnetism dominates over the iron one. So, from our analysis of the particular palladium-rich $Pd_{1-x}Fe_x$ alloy system we opt for the spin-fluctuation model, albeit not at the evidence level.

Analysis of literature on the iron concentration dependence of the Curie temperature [5,43] has shown that our results are in good agreement with that for the bulk alloys (see Fig. 6). This fact gives us further evidence of high compositional and structural uniformity of our epitaxial alloy films in spite of their relatively small thickness of 20 nm. The latter is a motivating factor for future study of the magnetic properties of $Pd_{1-x}Fe_x$ films with smaller thickness epitaxially grown following our procedures [31]. Reference [48], for example, reports on very strong dependence of $T_c$ on the film thickness for polycrystalline $Pd_{0.99}Fe_{0.01}$ alloy films prepared by magnetron sputtering.

*4.5. FMR measurements interpretation*

The angular dependence of the FMR field for resonance, Fig. 3, was analyzed using the free energy density expansion [49,50] assuming the tetragonal symmetry of the film lattice [32]:

$$E = -\mathbf{M} \cdot \mathbf{H} + 2\pi M_s^2 \alpha_3^2 - K_p \alpha_3^2 - \frac{1}{2} K_1 (\alpha_1^4 + \alpha_2^4 + \alpha_3^4) - \frac{1}{2} K_2 \alpha_3^4, \quad (5)$$

where the first term is the Zeeman energy in an external magnetic field; the second term describes the demagnetization energy arising from the thin-film shape of the sample; the third term, treated initially as the "perpendicular" anisotropy induced by the interface with the substrate and the free surface, finally accumulates also a leading term from the tetragonal distortion anisotropy of the same functional form as the perpendicular anisotropy is; the last two terms are the cubic and second-order tetragonal anisotropy contributions. The indexed $\alpha_i$'s represent the directional cosines of the magnetization vector $\mathbf{M}$ with respect to the crystallographic axes ([100], [010], [001]) of the MgO(001) substrate, and $M_s$ is a saturation magnetization at given temperature.

The anisotropy constants $K_i$ and spectroscopic $g$-factor are determined by simultaneous fitting of the in-plane (see Fig. 3) and out-of-plane angular dependencies of the resonance field. The model curves in Fig. 3 were obtained by simultaneous solution of the system of Suhl and equilibrium equations (see general prescriptions in the reviews [49,50], and for the particular case in Ref. [32]). Upon fitting the FMR data, the experimentally measured saturation magnetization $M_s$ was used to provide the stable convergence of the fitting procedure with the minimal scatter of the freely varied parameters.

The concentration dependences of the anisotropy energies and the spectroscopic $g$-factor are presented in Fig. 7. The figure shows that the magnitudes of the cubic, $K_1$, and the second-order tetragonal, $K_2$, anisotropy constants gradually with an increase of the iron concentration. The cumulative perpendicular and the first-order tetragonal anisotropy, $K_p$, reveals non-monotonous behavior: first, its positive magnitude increases, then it passes through a maximum at $x \sim 0.025$, changes sign at $x \sim 0.04$ and then steadily increases in the negative values range with large increment. To distinguish the evolution of the intrinsic anisotropies of $Pd_{1-x}Fe_x$ alloy from that associated with the thin film morphology we compare our FMR results with data found in the literature for the bulk sample and for various films. Indeed, FMR studies of the bulk and thick-film $Pd_{1-x}Fe_x$ [33,51-54] reported on only cubic anisotropy, the energy scale of which, $K_1$, is in a good agreement with ours (see Fig. 7). It is interesting that $K_1$ for nominally epitaxial and much thinner films of Ref. [27] also fit well our data, see Fig. 7. The absolute values of the anisotropy constants even for a sample with maximal iron concentration (- 26.6 kerg/cm$^3$ for $Pd_{0.92}Fe_{0.08}$) are quite small compared with the well-known from literature for epitaxial iron film $K_1$ = +480 kerg/cm$^3$ [55,56].

The $g$-factor for our films lies in the range of 2.13 – 2.2 (Fig. 7). It is a typical value for this material, for example, $g$ = 2.09 – 2.17 ($x$ = 0.025 – 0.06) [27,33,51,52,54]. The dependence of the $g$-factor on the iron content $x$ follows quite well the effective $g$-factor ($g_{eff}$) definition [33]:

$$g_{eff} = \frac{(1-x)\mu_{Pd} + c\mu_{Fe}}{(1-x)(\mu_{Pd}/g_{Pd}) + x(\mu_{Fe}/g_{Fe})} \quad (6)$$

where $\mu$ is magnetic moment of the particular kind of atom. The calculated dependence is shown in the bottom panel of Fig. 7 by the dashed line. Magnetic moment of 2.2$\mu_B$ for the Fe-atom was taken as the known value for pure iron [8,10]. Magnetic moment of the palladium atom was calculated from the values of low-temperature saturated magnetization of our films; it varies from 0.06$\mu_B$ to 0.20$\mu_B$ with an increase of iron content $x$ from 0.011 to 0.07. The values of the $g$-factors obtained from the fit were $g_{Pd}$ = 2.28 and $g_{Fe}$ = 2.0.

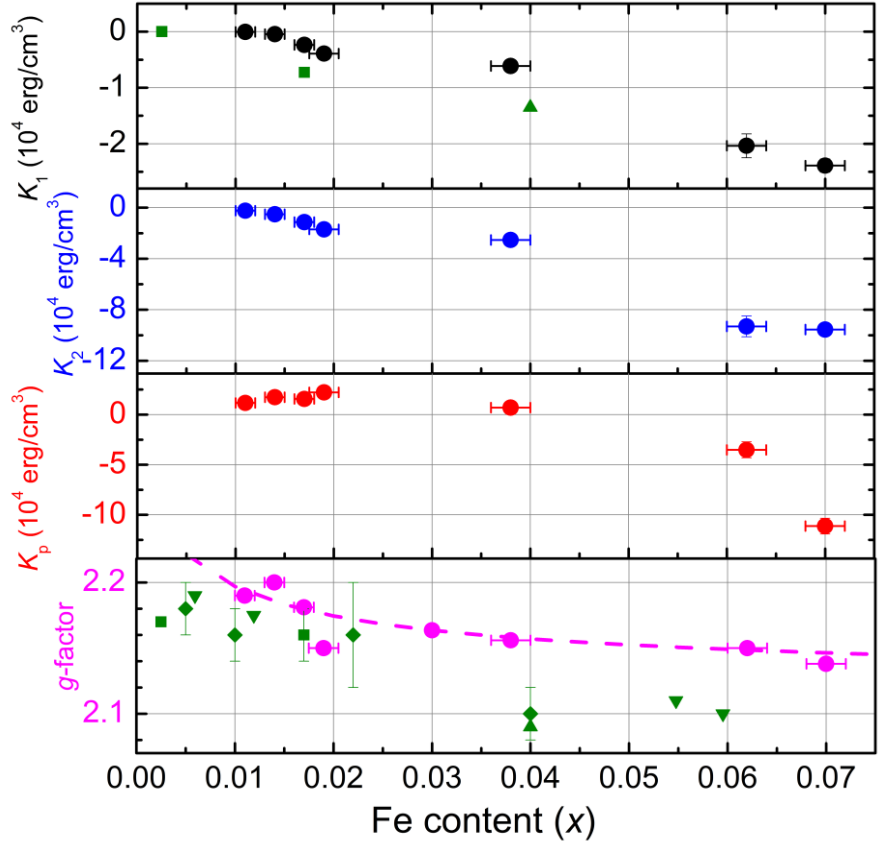

Figure 7. Dependences of the anisotropy constants $K_1$ (a), $K_2$ (b) and $K_p$ (c) and of the $g$-factor (circles, panel d). Dashed line in (d) presents the fit with Eq. (6). Results from other works are presented: for epitaxial film – ▲ [27], for bulk single crystal – ■ [33], for bulk polycrystal – ◆ [51,52], for a thick polycrystalline film – ▼ [54].

*4.5. Hysteresis loop analysis*

The FMR field for resonance in all studied films has values larger than 1000 Oe, which significantly exceeds both coercive fields (maximum 23 Oe, see Table 1) and saturation fields (several hundred Oe, see Fig. 4). Therefore, films under resonance conditions are inevitably in a saturated (single-domain) state. However, the desired properties of $Pd_{1-x}Fe_x$ thin films, mapped on their application in superconducting spintronics, are their magnetic uniformity and an ability of coherent rotation of the magnetization in the film plane by small magnetic fields. Such information can be obtained from an analysis of the shape of the hysteresis loops. Our epitaxial films are characterized by pronounced anisotropy in the film plane (Fig. 3), which is a prerequisite for a coherent (over the film volume) magnetization reversal. The most informative is the shape of the hysteresis loops in the field applied along the heavy in-plane axis; representative loops are shown in Fig. 4. It is seen that the shape of the loops is qualitatively identical for the three compositions presented.

To model the hysteresis loops we applied a model of the uniform magnetization rotation (Stoner-Wohlfarth mode). Equilibrium orientation of the magnetic moment $\varphi_M$ was determined by finding a minimum of the free energy, Eq. (5), at a given angle $\varphi_H$ and magnitude of the external magnetic field applied in the film plane. The anisotropy constant $K_1$ obtained from the analysis of FMR data has been used to model the hysteresis loops. The measured magnetization of the film corresponds to the projection of the equilibrium magnetization on the direction of the applied field [57].

It can be seen in Fig. 4 that the model loops reproduce well the shape observed in the experiment. Successful simulation of the hysteresis loops of thin epitaxial $Pd_{1-x}Fe_x$ films with $x \leq 0.07$ within the framework of the Stoner-Wohlfarth model allows us to state their predominant coherent magnetic moment rotation at low temperatures. This, in turn, confirms the promise of these materials for use as an F-layer in components of superconducting spintronics.

**Conclusions**

Epitaxial $Pd_{1-x}Fe_x$ ($0.011 \leq x \leq 0.07$) films were grown on single-crystal MgO (001) substrates, and magnetic properties of the films have been extensively studied with the combination of VSM magnetometry and ferromagnetic resonance measured in the in-plane and out-of-plane geometries. All fundamental magnetic parameters, such as the Curie temperature and saturation magnetization, as well as the magnetic anisotropy constants were determined, and their dependences on iron content $x$ were extracted. The shape of the magnetization dependence on temperature was found universal in the studied iron content range. It is extensively discussed in the framework of several models of dilute magnetic alloys, and it was concluded that the spin-fluctuation model has the least controversies with our experimental data. The in-plane magnetic hysteresis loops were successfully reproduced utilizing the FMR data for the magnetic anisotropies and the Stoner-Wohlfarth model of the magnetic moment reversal. This allowed us to state the predominant coherent magnetic moment rotation upon remagnetization at low temperatures. The presented experimental data and their analysis could be useful in constructing elements of cryogenic spintronics and quantum computing circuits based on Josephson π-contacts.

**Acknowledgements**

The support of the work by RSF grant No. 18-12-00459 is gratefully acknowledged. The authors indebted to V.V. Ryazanov, V.V. Bol'ginov and V.S. Stolyarov (ISSP of RAS, Chernogolovka) for stimulating discussions and introduction to the subject. TEM study was done using equipment of "Physics and technology of micro- and nanostructures" Center at IPM RAS, Nizhny Novgorod. The magnetic properties were measured utilizing the equipment of the PCR Federal Center of Shared Facilities of Kazan Federal University.

**References**

1. D. Gerstenberg, Magnetische Untersuchungen an Pd-Mischkristallen mit Übergangselementen. *Ann. Phys. (Leipzig)* 2 (1958) 236-262.

2. J. Crangle, Ferromagnetism in Pd-rich palladium-iron alloys. *Phil. Mag.* 5 (1960) 335-342.

3. J. Crangle and W.R. Scott, Dilute ferromagnetic alloys. *J. Appl. Phys.* 36 (1965) 921-928.

4. A.M. Clogston, B.T. Matthias, M. Peter, H.J. Williams, E. Corenzwit, and R.C. Sherwood, Local magnetic moment associated with an iron atom dissolved in various transition metal alloys. *Phys. Rev.* 125 (1962) 541-552.

5. G.J. Nieuwenhuys, Magnetic behaviour of cobalt, iron and manganese dissolved in palladium. *Advances in Physics* 24 (1975) 515-591.

6. K. Yosida, Theory of Magnetism, *Springer Series in Solid-State Sciences* v.122, Springer 1996, Part III.

7. K.H.J. Buschow and F.R. de Boer, Physics of Magnetism and Magnetic Materials, *Kluwer Academic Publishers* 2004, Ch. 7.

8. P. Mohn, Magnetism in the Solid State, *Springer Series in Solid-State Sciences* v.134, Springer 2006, Ch. 8.

9. M. Getzlaff, Fundamentals of Magnetism, *Springer* 2008, Ch. 3.

10. J.M.D. Coey, Magnetism and Magnetic Materials, *Cambridge University Press* 2010, Ch. 5.

11. T.I. Larkin, V.V. Bol'ginov, V.S. Stolyarov, V.V. Ryazanov, I.V. Vernik, S.K. Tolpygo, O.A. Mukhanov, Ferromagnetic Josephson switching device with high characteristic voltage. *Appl. Phys. Lett.* 100 (2012) 222601.

12. I.V. Vernik, V.V. Bol'ginov, S.V. Bakurskiy, A.A. Golubov, M.Yu. Kupriyanov, V.V. Ryazanov, and O. Mukhanov, Magnetic Josephson junctions with superconducting interlayer for cryogenic memory. *IEEE Trans. Appl. Supercond.* 23 (2013) 1701208.

13. B.M. Niedzielski, S.G. Diesch, E.C. Gingrich, Y. Wang, R. Loloee, W.P. Pratt, Jr., and N.O. Birge. Use of Pd–Fe and Ni–Fe–Nb as Soft Magnetic Layers in Ferromagnetic Josephson Junctions for Nonvolatile Cryogenic Memory. *IEEE Trans. Appl. Supercond.* 24 (2014) 1800307.

14. M.A. Manheimer, Cryogenic Computing Complexity Program: Phase 1. Introduction. *IEEE Trans. Appl. Supercond.* 25 (2015) 1301704.

15. I.A. Golovchanskiy, V.V. Bol'ginov, N.N. Abramov, V.S. Stolyarov, A. Ben Hamida, V.I. Chichkov, D. Roditchev, and V.V. Ryazanov, Magnetization dynamics in dilute $Pd_{1-x}Fe_x$ thin films and patterned microstructures considered for superconducting electronics. *J. Appl. Phys.* 120 (2016) 163902.

16. V.V. Ryazanov, Superconductor-ferromagnet-superconductor Josephson pi-junction as an element of a quantum bit (experiment). *Phys.-Usp.* 42 (1999) 825–827.

17. L. Lazar, K. Westerholt, H. Zabel L.R. Tagirov Yu.V. Goryunov, N.N. Garif'yanov, I.A. Garifullin. *Phys. Rev. B* 61 (2000) 3711-3722.

18. L.R. Tagirov, I.A. Garifullin, N.N. Garifyanov, S.Ya. Khlebnikov, D.A. Tikhonov, K. Westerholt, H. Zabel. *J. Magn. Magn. Mater.* 240 (2002) 577-579.

19. A.S. Sidorenko, V.I. Zdravkov, A.A. Prepelitsa, C. Helbig, Y. Luo, S. Gsell, M. Schreck, S. Klimm, S. Horn, L.R. Tagirov, R. Tidecks. *Ann. Phys. (Leipzig)* 12 (2003) 37–50.

20. A.I. Buzdin, Proximity effects in superconductor-ferromagnet heterostructures. *Rev. Mod. Phys.* 77 (2005) 935-976.

21. D.J. Webb and J.D. McKinley Two-dimensional magnetism in Pd (1.2 at.% Fe) films. *Phys. Rev. Lett.* 70 (1993) 509-511.

22. H.Z. Arham, T.S. Khaire, R. Loloee, W.P. Pratt Jr, N.O. Birge, Measurement of spin memory lengths in PdNi and PdFe ferromagnetic alloys. *Phys. Rev. B* 80 (2009) 174515.

23. L.S. Uspenskaya, A.L. Rakhmanov, L.A. Dorosinskii, A.A. Chugunov, V.S. Stolyarov, O.V. Skryabina, S.V. Egorov. Magnetic patterns and flux pinning in $Pd_{0.99}Fe_{0.01}$-Nb hybrid structures. *JETP Letters* 97 (2013) 155-158.

24. L.S. Uspenskaya, A.L. Rakhmanov, L.A. Dorosinskii, S.I. Bozhko, V.S. Stolyarov, V.V. Bol'ginov. Magnetism of ultrathin $Pd_{99}Fe_{01}$ films grown on niobium. *Materials Research Express* 1 (2014) 036104.

25. V.V. Bol'ginov, O.A. Tikhomirov, L.S. Uspenskaya, Two-component magnetization in $Pd_{99}Fe_{01}$ thin films. *JETP Letters* 105 (2017) 169-173.

26. M. Schöck, C. Sürgers, H.V. Löhneysen, Superconducting and magnetic properties of $Nb/Pd_{1-x}Fe_x/Nb$ triple layers. *Eur. Phys. Journ. B* 14 (2000) 1-10.

27. I.A. Garifullin, , D.A. Tikhonov, N.N. Garif'yanov, M.Z. Fattakhov, K. Theis-Bröhl, K. Westerholt, H. Zabel, Possible reconstruction of the ferromagnetic state under the influence of superconductivity in epitaxial $V/Pd_{1-x}Fe_x$ bilayers. *Appl. Magn. Resonance* 22 (2002) 439-452.

28. R.I. Salikhov, N.N. Garif'yanov, I.A. Garifullin, L.R. Tagirov, K. Westerholt, H. Zabel, Spin screening effect in superconductor/ferromagnet thin film heterostructures studied using nuclear magnetic resonance. *Phys. Rev. B* 80 (2009) 2145231.

29. M. Ewerlin, B. Pfau, C.M. Günther, S. Schaffert, S. Eisebitt, R. Abrudan, H. Zabel, Exploration of magnetic fluctuations in PdFe films. *Journ. Phys.: Cond. Matter*, 25 (2013) 266001.

30. T. Wagner, G. Richter, M. Rühle, Epitaxy of Pd thin films on (100) $SrTiO_3$: A three-step growth process. *J. Appl. Phys.* 89 (2001) 2606-2612.

31. A. Esmaeili, I.V. Yanilkin, A.I. Gumarov, I.R. Vakhitov, B.F. Gabbasov, A.G. Kiiamov, A.M. Rogov, Yu.N. Osin, A.E. Denisov, R.V. Yusupov, L.R. Tagirov. Epitaxial growth of thin $Pd_{1-x}Fe_x$ films on MgO single crystal. *Thin Solid Films* 669 (2019) 338-344.

32. A. Esmaeili, I.R. Vakhitov, I.V. Yanilkin, A.I. Gumarov, B.M. Khaliulin, B.F. Gabbasov, M.N. Aliyev, R.V. Yusupov, L.R. Tagirov. FMR studies of ultra-thin epitaxial $Pd_{0.92}Fe_{0.08}$ film. *Appl. Magn. Resonance* 49 (2018) 175–183.

33. D.M.S. Bagguley_1, J.A. Robertson, Resonance and magnetic anisotropy in dilute alloys of Pd, Pt with Fe, Co and Ni. *Journ. Phys. F: Metal Phys.* 4 (1974) 2282-2296.

34. K.K. Murata, S. Doniach. Theory of Magnetic Fluctuations in Itinerant Ferromagnets, *Phys. Rev. Lett.* 29 (1972) 285-288.

35. D. Bloch, D.M. Edwards, M. Shimizu, J. Voiron, First order transitions in $ACo_2$ compounds, *J. Phys. F: Metal Phys.* 5 (1975) 1217-1226.

36. G.G. Lonzarich, L. Taillefer. Effect of spin fluctuations on the magnetic equation of state of ferromagnetic or nearly ferromagnetic metals, *J. Phys. C: Solid State Phys.* 18 (1985) 4339-4371.

37. D. Wagner, The fixed-spin-moment method and fluctuations, *J. Phys.: Condens. Matter* 1 (1989) 4635-4642.

38. M.D. Kuz'min, Landau-type parametrization of the equation of state of a ferromagnet. *Phys. Rev. B* 77 (2008) 184431.

39. A. Waske, H. Hermann, N. Mattern, K. Skokov, O. Gutfleisch, J. Eckert, Magnetocaloric effect of an Fe-based metallic glass compared to benchmark gadolinium. *Journ. Appl. Phys.* 112 (2012) 123918.

40. K. Handrich, A simple model for amorphous and liquid ferromagnets. *Phys. Stat. Sol. (b)* 32 (1969) K55-K58.

41. S. Kobe. Spontaneous Magnetization of an Amorphous Ferromagnet, *Phys. Stat. Sol. (b)* 41 (1970) K13-K15.

42. K.A. Gallagher, M.A. Willard, V.N. Zabenkin, D.E. Laughlin, M.E. McHenry. Distributed exchange interactions and temperature dependent magnetization in amorphous $Fe_{88-x}Co_xZr_7B_4Cu_1$ alloys. *Journ. Appl. Phys.* 85 (1999) 5130-5132.

43. J.C. Ododo, Percolation concentration and saturation of the Pd moment in ferromagnetic Pd alloys. *Journal of Physics F: Metal Physics*, 13 (1983) 1291-1309.

44. J.C. Ododo, Ferromagnetic correlation lengths in dilute PdFe and PdCo alloys. *Journal of Physics F: Metal Physics*, 15 (1985) 941-951.

45. I.Ya. Korenblit, E.F. Shender, Ferromagnetism of disordered systems. *Soviet Phys. - Uspekhi* 21 (1978) 832-851.

46. D.F. Holcomb, J.J. Rehr, Jr, Percolation in heavily doped semiconductors. *Phys. Rev.* 183 (1969) 773-776.

47. P. Mohn, K. Schwarz, Supercell calculations for transition metal impurities in palladium. *J. Phys.: Condens. Matter* 5 (1993) 5099-5112.

48. V.V. Bol'ginov, V.S. Stolyarov, D.S. Sobanin, A.L. Karpovich, V.V. Ryazanov, Magnetic switches based on Nb-PdFe-Nb Josephson junctions with a magnetically soft ferromagnetic interlayer. *JETP Letters* 95 (2012) 366-371.

49. B. Heinrich, J.F. Cochran, Ultrathin metallic magnetic films: magnetic anisotropies and exchange interactions. *Adv. Phys.* 42, (1993) 523-639.

50. M. Farle, Ferromagnetic resonance of ultrathin metallic layers. *Rep. Progr. Phys.* 61 (1998) 755-826.

51. D.M.S. Bagguley, W.A. Crossley, J. Liesegang, Ferromagnetic resonance in a series of alloys: II. Binary alloys of cobalt with platinum and palladium, and one iron-palladium alloy. *Proc. Phys. Soc.* 90 (1967) 1047-1058.

52. D.M.S. Bagguley, J.A. Robertson, Ferromagnetic resonance in dilute binary alloys of Pd and Pt with Fe and Co. *Phys. Lett. A* 27 (1968) 516-517.

53. N. Miyata, K. Tomotsune, H. Nakada, M. Hagiwara, H. Kadomatsu, H. Fujiwara, Ferromagnetic Crystalline Anisotropy of $Pd_{1-x}Fe_x$ Alloys. I ($x \lesssim 0.3$, FCC Phase). *Journ. Phys. Soc. Japan*, 55 (1986) 946-952.

54. D.L. Hardison, E.D. Thompson, Spin wave resonance on PdFe alloys, Journ. Phys. Colloq. 32(C1) (1971) 565-566.

55. Yu.V. Goryunov, N.N. Garif'yanov, G.G. Khaliullin, I.A. Garifullin, L.R. Tagirov, F. Schreiber, Th. Muhge, H. Zabel. Magnetic anisotropies of sputtered Fe films on MgO substrates, Phys. Rev. B 52 (1995) 13450-13458.

56. N. Tournerie, P. Schieffer, B. Lépine, C. Lallaizon, P. Turban, G. Jézéquel. In-plane magnetic anisotropies in epitaxial Fe(001) thin films. *Phys. Rev. B* 78 (2008) 134401.

57. S.V. Vonsovskii, E.R. Hardin, *Magnetism* (Vol. 2, 1974), New York: Wiley, 635 p.